\documentclass[12pt]{article}
\usepackage{mathrsfs}
\usepackage{amssymb}
\usepackage{amsfonts}
\usepackage{latexsym}
\usepackage{amsthm}

\theoremstyle{plain}
\newtheorem{theorem}{\bf Theorem}[section]
\newtheorem{lemma}[theorem]{\bf Lemma}

\theoremstyle{remark}

\newcommand{\er}[1]{{\rm(\ref{#1})}}
\def\lb{\label}

\begin{document}
\def\a{\alpha}  \def\mA{{\mAl A}}    \def\bA{{\bf A}}  \def\mA{{\mathscr A}}
\def\b{\beta}   \def\cB{{\mAl B}}    \def\bB{{\bf B}}  \def\mB{{\mathscr B}}
\def\g{\gamma}  \def\cC{{\mAl C}}    \def\bC{{\bf C}}  \def\mC{{\mathscr C}}
\def\G{\Gamma}  \def\cD{{\mAl D}}    \def\bD{{\bf D}}  \def\mD{{\mathscr D}}
\def\d{\delta}  \def\cE{{\mAl E}}    \def\bE{{\bf E}}  \def\mE{{\mathscr E}}
\def\D{\Delta}  \def\cF{{\mAl F}}    \def\bF{{\bf F}}  \def\mF{{\mathscr F}}
\def\c{\chi}    \def\cG{{\mAl G}}    \def\bG{{\bf G}}  \def\mG{{\mathscr G}}
\def\z{\zeta}   \def\cH{{\mAl H}}    \def\bH{{\bf H}}  \def\mH{{\mathscr H}}
\def\e{\eta}    \def\cI{{\mAl I}}    \def\bI{{\bf I}}  \def\mI{{\mathscr I}}
\def\p{\psi}    \def\cJ{{\mAl J}}    \def\bJ{{\bf J}}  \def\mJ{{\mathscr J}}
\def\vT{\Theta} \def\cK{{\mAl K}}    \def\bK{{\bf K}}  \def\mK{{\mathscr K}}
\def\k{\kappa}  \def\cL{{\mAl L}}    \def\bL{{\bf L}}  \def\mL{{\mathscr L}}
\def\l{\lambda} \def\cM{{\mAl M}}    \def\bM{{\bf M}}  \def\mM{{\mathscr M}}
\def\L{\Lambda} \def\cN{{\mAl N}}    \def\bN{{\bf N}}  \def\mN{{\mathscr N}}
\def\m{\mu}     \def\cO{{\mAl O}}    \def\bO{{\bf O}}  \def\mO{{\mathscr O}}
\def\n{\nu}     \def\cP{{\mAl P}}    \def\bP{{\bf P}}  \def\mP{{\mathscr P}}
\def\r{\rho}    \def\cQ{{\mAl Q}}    \def\bQ{{\bf Q}}  \def\mQ{{\mathscr Q}}
\def\s{\sigma}  \def\cR{{\mAl R}}    \def\bR{{\bf R}}  \def\mR{{\mathscr R}}
\def\S{\Sigma}  \def\cS{{\mAl S}}    \def\bS{{\bf S}}  \def\mS{{\mathscr S}}
\def\t{\tau}    \def\cT{{\mAl T}}    \def\bT{{\bf T}}  \def\mT{{\mathscr T}}
\def\f{\phi}    \def\cU{{\mAl U}}    \def\bU{{\bf U}}  \def\mU{{\mathscr U}}
\def\F{\Phi}    \def\cV{{\mAl V}}    \def\bV{{\bf V}}  \def\mV{{\mathscr V}}
\def\P{\Psi}    \def\cW{{\mAl W}}    \def\bW{{\bf W}}  \def\mW{{\mathscr W}}
\def\o{\omega}  \def\cX{{\mAl X}}    \def\bX{{\bf X}}  \def\mX{{\mathscr X}}
\def\x{\xi}     \def\cY{{\mAl Y}}    \def\bY{{\bf Y}}  \def\mY{{\mathscr Y}}
\def\X{\Xi}     \def\cZ{{\mAl Z}}    \def\bZ{{\bf Z}}  \def\mZ{{\mathscr Z}}
\def\O{\Omega}
\def\ve{\varepsilon}
\def\eps{\epsilon}
\def\vt{\vartheta}
\def\vp{\varphi}
\def\vk{\varkappa}

\def\Z{{\Bbb Z}}
\def\R{{\Bbb R}}
\def\C{{\Bbb C}}
\def\T{{\Bbb T}}
\def\N{{\Bbb N}}

\def\qqq{\qquad}
\def\qq{\quad}
\let\ge\geqslant
\let\le\leqslant
\def\ma{\left(\begin{array}{cc}}
\def\am{\end{array}\right)}
\def\iint{\int\!\!\!\int}
\def\lt{\biggl}
\def\rt{\biggr}
\let\geq\geqslant
\let\leq\leqslant
\def\[{\begin{equation}}
\def\]{\end{equation}}
\def\wt{\widetilde}
\def\pa{\partial}
\def\sm{\setminus}
\def\es{\emptyset}
\def\no{\noindent}
\def\ol{\overline}
\def\iy{\infty}
\def\ev{\equiv}
\def\/{\over}
\def\ts{\times}
\def\os{\oplus}
\def\ss{\subset}
\def\h{\hat}
\def\Re{\mathop{\rm Re}\nolimits}
\def\Im{\mathop{\rm Im}\nolimits}
\def\supp{\mathop{\rm supp}\nolimits}
\def\sign{\mathop{\rm sign}\nolimits}
\def\Ran{\mathop{\rm Ran}\nolimits}
\def\Ker{\mathop{\rm Ker}\nolimits}
\def\Tr{\mathop{\rm Tr}\nolimits}
\def\const{\mathop{\rm const}\nolimits}
\def\Wr{\mathop{\rm Wr}\nolimits}
\def\Ai{\mathop{\rm Ai}\nolimits}
\def\Bi{\mathop{\rm Bi}\nolimits}
\def\BBox{\hspace{1mm}\vrule height6pt width5.5pt depth0pt \hspace{6pt}}

\def\Twelve{
\font\Tenmsa=msam10 scaled 1200
\font\Sevenmsa=msam7 scaled 1200
\font\Fivemsa=msam5 scaled 1200
%\newfam\msafam
\textfont\msafam=\Tenmsa
\scriptfont\msafam=\Sevenmsa
\scriptscriptfont\msafam=\Fivemsa
\font\Tenmsb=msbm10 scaled 1200
\font\Sevenmsb=msbm7 scaled 1200
\font\Fivemsb=msbm5 scaled 1200
%\newfam\msafam
\textfont\msbfam=\Tenmsb
\scriptfont\msbfam=\Sevenmsb
\scriptscriptfont\msbfam=\Fivemsb

\font\Teneufm=eufm10 scaled 1200
\font\Seveneufm=eufm7 scaled 1200
\font\Fiveeufm=eufm5 scaled 1200
%\newfam\eufmfam
\textfont\eufmfam=\Teneufm
\scriptfont\eufmfam=\Seveneufm
\scriptscriptfont\eufmfam=\Fiveeufm}

\def\Ten{
\textfont\msafam=\tenmsa
\scriptfont\msafam=\sevenmsa
\scriptscriptfont\msafam=\fivemsa

\textfont\msbfam=\tenmsb
\scriptfont\msbfam=\sevenmsb
\scriptscriptfont\msbfam=\fivemsb

\textfont\eufmfam=\teneufm
\scriptfont\eufmfam=\seveneufm
\scriptscriptfont\eufmfam=\fiveeufm}

\title {Marchenko-Ostrovski mappings for periodic Jacobi matrices}

\author{ Evgeny Korotyaev
\begin{footnote}
{Institut f\"ur  Mathematik,  Humboldt Universit\"at zu Berlin,
Rudower Chaussee 25, 12489, Berlin, Germany, e-mail:
evgeny@math.hu-berlin.de }
\end{footnote}
and  Anton Kutsenko${}^{\ast,\!\!\!}$
\begin{footnote}
{Faculty of Math. and Mech. St-Petersburg State University, e-mail:
kucenkoa@rambler.ru}
\end{footnote}
} \maketitle

\begin{abstract}
\no We consider the 1D periodic Jacobi matrices.
The spectrum of this operator is purely absolutely continuous and  consists of intervals separated by gaps. We solve the inverse problem (including characterization) in terms of vertical slits on the quasimomentum domain .
Furthermore, we obtain a priori  two-sided estimates for 
vertical slits in terms of Jacoby matrices.
\end{abstract}

\vskip 0.25cm
\section {Introduction and main results}
\setcounter{equation}{0}

Consider the Jacobi matrices $L$ on $l^2(\Z)$ given by
 $(Ly)_n=a_{n-1}y_{n-1}+a_{n}y_{n+1}+b_ny_n,\ n\in\Z$,
where $a_n=e^{x_n}>0, x_n, b_n\in\R$ are $N-$periodic sequences and
let
 \[
\lb{1a} 
x=(x_n)_1^N,\  b=(b_n)_1^N,\ \  p=(x,b)\in\mH^2,\
\ \ \mH\ev\lt\{b\in\R^{N}:\ \sum_1^Nb_n=0\rt\}.
\]
since we can take the number $a_n$ such that $a_1a_2..a_N=1$
(after the multiplication on some number).
 Introduce fundamental solutions
$\vp=(\vp_n(\l,p))_{n\in\Z}$ and $ \vt=(\vt_n(\l,p))_{n\in \Z}
$ of the equation
\[
 \lb{1e}
a_{n-1} y_{n-1}+a_{n}y_{n+1}+b_ny_n=\l y_n,\ \ (\l,n)\in\C\ts\Z,
\]
with initial conditions $\vp_{0}\ev \vt_1\ev 0,\ \vp_1\ev
\vt_{0}\ev 1$.  The function $\D(\l,p)=\vp_{N+1}(\l,p)+\vt_N(\l,p)$ is called the Lyapunov function for the operator $L$. The functions $\D, \vp_n$ and $\vt_n, n\ge1$ are polynomials of $(\l,a,b)\in\C^{2N+1}$.
It is well known [vM] that the spectrum of $L$ is absolutely continuous
and consists of $N$ intervals $\s_n=[\l_{n-1}^+,\l_{n}^-],
\ n\in \N_N=\{1, ... ,N\}$, where $\l_n^\pm =\l_n^\pm (p)$ and
$\l_{N}^+\ev\l_0^{+}<\l_1^-\le \l_1^+< ... < \l_{N-1}^- \le \l_{N-1}^+
< \l_{N}^{-} $. These intervals are separated by gaps
$\g_n=(\l_n^-,\l_n^+)$ of lengths $|\g_n|\ge 0$. If a gap
$\g_n$ is degenerate, i.e. $|\g_n|=0$, then the corresponding
segments $\s_n$, $\s_{n+1}$ merge. The spectrum of $L$ is given by
$\s(p)=\{\l\in\R:\ |\D(\l,p)|\le 2\}$ and note that
$(-1)^{N-n}\D(\l_n^{\pm},p)=2, n\in \N_N$.  For each $n\in \N_ {N-1}$
there exists unique $\l_n=\l_n(p)\in[\l_n^-,\l_n^+]$  such that
\[
\lb{1c}
 \D'(\l_n,p)=0,\ \ \D''(\l_n,p)\neq 0,\ \
 (-1)^{s_n}\D(\l_n,p)\ge2, \ \ s_n=N-n.
\]
Here and below we use the notation $(\ ')={\pa/\pa\l}$. Let
$\m_n=\m_n(p)$ be the zeros of $\vt_{N+1}(\l,p)$. It is well
known that $\m_n\in[\l_n^-,\l_n^+], n\in \N_ {N-1}$. Define the
Marchenko-Ostrovski mapping $h:\mH^2\to \R^{2N-2}$ by
$h(p)=(h_n(p))_{1}^{N-1}$, where the components
$h_n=(h_{1n},h_{2n})\in \R^2$ and
\[
\lb{1d}  \ \ \ h_{1n}=\log [(-1)^{s_n}\vt_{N}(\m_n)], \ \
 \ \ h_{2n}=||h_n|^2-h_{1n}^2|^{1/2}{\rm sign} (\l_n-\m_n).
\]
Note that $(-1)^{s_n}\vt_{N}(\m_n)>0$. Here the function
$|h_n(p)|^2, p\in \mH^2,$ is defined by
\[
\lb{a7} 2\cosh |h_n|=(-1)^{s_n}\D(\l_n(p),p),\ \ \ \ p\in \mH^2.
\]
The Wronskian identity  $\vt_{N}\vp_{N+1}-\vt_{N+1}\vp_{N}=1$ and
\er{1d} imply
\[
\lb{i}
 \vp_{N+1}(\m_n,p)\vt_{N}(\m_n,p)=1,\ \ \ \ \ \
 (-1)^{s_n}\D(\m_n(p),p)=2\cosh h_{1n},\ \ \  n\in \N_ {N-1}.
\]
Note that \er{i},\er{a7} gives $|h_n|^2-h_{1n}^2\ge 0,$ since
$(-1)^{s_n}\D$ has the maximum at $\l_n$ on the segment
$[\l_n^-,\l_n^+]$. We formulate our main result.

 \begin{theorem} \lb{T1}
The mapping $h:\mH^2\to \R^{2N-2}$ is a real analytic isomorphism
between $\mH^2$ and $\R^{2N-2}$. Moreover, the following estimates
hold (here $c={\l_N^--\l_0^+\/2}$)
\[
\lb{e} (1/4)e^{2h_+}<c^2<\sum_1^N(b_n^2+2a_n^2)<4Nc^2<32Ne^{2h_+},\ \ \ \
h_+\ev\max |h_n|,\ a=(a_n)_1^N.
\]
\end{theorem}

\no{\bf Remark.} The vector $a=(a_n)_1^N$ belongs to the
manifold $\mA\ev\lt\{a\in\R_+^{N}:\prod_1^Na_n=1\rt\}\ss \R^N$.
The map $\r:\mA\to \mH$ given by $\r(a)=x, x_n=\log a_n$, is a
real analytic isomorphism between $\mA$ and $\mH$. Then by Theorem
1, the map $h\circ (\r\ts I):\mA\ts\mH\to \R^{2N-2}$ is a real
analytic isomorphism between $\mA\ts\mH$ and $\R^{2N-2}$.

 The mapping $h$ is some analog of the Marchenko and Ostrovski mapping  for the continuous case [MO]
 and $h$  has the similar geometric interpretation in terms of conformal mapping (see [MO], [K1]). We extend the result
of  Marchenko and Ostrovski  about the height-slit mapping for the Hill operator (see [MO], [K], [K1]) to the case of the periodic Jacobi matrix. 
We describe the geometric sense of the map $h$. Introduce a domain
$\L=\C\sm\cup_1^{N-1}\g_n$ and a quasimomentum domain $
K=\{k:0\le\Re k\le N\pi\}\sm\cup_1^{N-1}\G_n,\ \ \G_n =(\pi
n+i|h_n|,\pi n-i|h_n|)$.

\no {\bf Corollary 1.2.}
 {\it For each $h\in \R^{2N-2}$ there exist a unique
$p\in \mH^2$ and a unique conformal mapping $k:\L\to K$ such that
following identities and asymptotics hold}:
\[
\lb{ca}
  2\cos k(\l)=(-1)^N\D (\l,p),   \ \ \ \l\in \L, \ \ \ \ \ {\rm and} \ \ \
  \ \ \ k(it)\to \pm i \iy\ \ \ {\rm as} \ \ t\to \pm\iy,
\]
\[
\lb{ci} 
k(\l_n(p)\pm i0)=\pi n\pm i|h_n|,   \ \ \ k(\m_n(p)\pm
i0)=\pi n\pm ih_{1n},   \ \ \ n\in \N_ {N-1}.
\]
{\bf Remark about recovering.} The function $k:\L\to K$ has
the properties: $k(\s_n)=[\pi (n-1),\pi n], n\in \N_ {N}$ and $k(\g_n)=\G_n,
n\in \N_ {N-1}$. Thus if we know $(|h_n|)_1^{N-1}$, then we obtain
$\L$. Furthermore, \er{ci} gives all $\m_n, n\in \N_{N-1}$.
Then the standard algorithm  for
$\m_n, h_{1n}, n\in \N_{N-1}$ determines $a,b$, see p.50, [vM] 
where reconstruction of $(a,b)$ in terms of spectral data
was treated already earlier.

\no {\bf Proof.} By Theorem 1.1, for each $h\in \R^{2N-2}$ there exists
unique $p\in \mH^2$ such that identities \er{1d}-\er{i} hold true.
Moreover, for any $p\in \mH^2$ there exists a unique conformal
mapping $k:\L\to K $ with the properties \er{ca} (see [KoK]),
which together with \er{1d}-\er{i} yields \er{ci}. $\BBox$

In order to prove Theorem \ref{T1} we use the direct approach from
[KK], based on a theorem from nonlinear functional analysis. We
improve a "basic theorem" of the direct method.

\no   {\bf Theorem A.} {\it Let $\mH, \mH_0$ be Hilbert spaces
equipped with norms $\|\cdot \|, \|\cdot \|_0$. Let a map $f: \mH
\to \mH_0$ satisfy conditions:

\no  i) $f$ is a local homeomorphism,

\no  ii) $f-f_0$ maps a weakly
convergent sequence in $\mH$ into a  strongly convergent sequence
in $\mH_0$, where $f_0: \mH \to \mH_0$ is a
homeomorphism between $\mH$ and $\mH_0$,

\no iii) $\|f(x)\|_0\to \iy $ as $\|x\|\to \iy$  and
$f^{-1}(0)=0$.

\no  Then $f$ is a homeomorphism between $\mH$ and $\mH_0$.}

{\bf Remark.} We recall definitions. Let $\mH, \mH_0$ be Hilbert
spaces.
% equipped with norms $\|\cdot \|, \|\cdot \|_0$.
The derivative of a map
$f:\mH\to \mH_0$ at a point $p\in \mH$ is a bounded linear
 map from $\mH$ into $\mH_0$, which we denote by $d_pf$.
 A map $f:\mH\to\mH_0$ is a real
analytic isomorphism between $\mH$ and $\mH_0$
 if $f$ is one-to-one and onto
and both $f$ and $f^{-1}$ are real analytic maps of the space. Let
$f$ satisfy all conditions in Theorem A and $f$ be real analytic (
or of class $C^s, s\ge 1$), and let the operator $d_pf$ have an
inverse for all $p \in \mH$. Then Theorem A and the Inverse
Function Theorem yield that $f$ is a real analytic (respectively,
$C^s$-) isomorphism between $\mH, \mH_0$.

There are various methods of solving inverse problems for periodic
potentials, see [MO], [GT],[vM], [KK], [K]. 
Recently, the author [K1] extended the results of [MO], [GT], [K] for the case $-y''+uy $ to the case of distributions, i.e. $-y''+u'y$ on $L^2(\R)$, where periodic $u\in L_{loc}^2(\R)$.
We know only three papers about the characterization of the
spectrum of periodic Jacobi matrices [Pe], [BGGK], [K2].
 Following the Marchenko-Ostrovski approach [MO], Perkolab [Pe] obtained characterization of the spectrum of periodic Jacobi matrices,
 but he did not show that $h$ is a homeomorphism between $\mH^2$ and $\R^{2N-2}$.
In fact, he proved some analog of Corollary 1.2. In [BGGK], [K2] the
inverse problems in terms of gap lengths  were solved. 

 The plan of the paper is as follows.
Firstly, we prove Theorem A. Secondly  we verify conditions
i)-iii) of Theorem A for the mapping $h$ and here we essentially
use the paper [K]. The analyticity of $h_{1n}$ is a simple fact.
The main problem is analyticity of $h_{2n}$ and Lemma \ref{L23} is
crucial. To check ii), we prove that each Frech\'et derivative
$d_ph,p\in \mH^2$  is invertible. We assume that there
exists $g\in \mH^2, g\neq 0$ such that $(d_ph)g=0$. We define the
polynomial $f(\l)\ev<(\pa \D)(\l,p),g>,  \l\in \C$,
 of degree $N-2$ with respect to $\l$ ($<\cdot,\cdot>$ is
 the inner product in $\R^{2N}$) and show that $f\ev0$.
 Using the last fact and the
result that $\{d_ph_{1n},d_p\m_{n}\}_1^{N-1}$ is a basis of
$\mH^2$ [vM], we get that $\{d_ph_{1n},d_ph_{2n}\}_1^{N-1}$ is a
basis of $\mH^2$, which yields $g=0$, i.e., we have a
contradiction. The verification of iii) is based on the estimates
from [KoK].

The motivation of this paper is to study the inverse problem for
the Schr\"odinger operator, i.e., $a_n=1, n\in \Z$. Note that a
characterization for this case is absent, see [KKu].

\section {Proof}
\setcounter{equation}{0}

\no   {\bf Proof of  Theorem A.} Using Conditions i), we see that
the set $f(\mH)$ is open. We prove that it is also closed. Suppose
that $h_n=f(q_n)\to h$ strongly as $n\to \iy $ for some $q_n\in
\mH$. Then Condition iii) yields $\sup \|q_n\|<\iy$. Hence there
exists a subsequence $\{q_{n_m}\}_{m=1}^{\iy}$ such that
$q_{n_m}\to q$ weakly as ${m\to\iy }$. Therefore, for $K=f-f_0$
Condition ii) implies $h_{n_m}-K(q_{n_m})=f_0(q_{n_m})\to h-K(q)$
as $m\to \iy$. Then $q_{n_m}\to q$ strongly as $m\to \iy$, since
$f_0$ is a homeomorphism between $\mH$ and $\mH_0$. Thus $f(\mH)$
is closed.

We show that $f$ is an injection. We introduce the set $S=\{q\in
\mH: f(q)=f(p)$ for some $q\neq p\in \mH\}$. We will show that
$S=\es$. Firstly, $S$ is open, since $f$ is a local isomorphism.

Secondly we prove that $S$ is closed. Suppose that $q_n\to q$
strongly as $n\to \iy $ and $f(q_n)=f(p_n)$ for some $q_n,p_n\in
S$ and $q_n\neq p_n$. Then $h_n=f(q_n)=f(p_n)\to h$ strongly as
$n\to \iy $. Then Condition iii) yields $\sup \|p_n\|<\iy$.
 Hence there exists a subsequence $\{p_{n_m}\}_{m=1}^{\iy}$ such
that $p_{n_m}\to p$ weakly as ${m\to\iy }$. Therefore, Condition
ii) implies $h_{n_m}-K(p_{n_m})=f_0(p_{n_m})\to h-K(p)$ as $m\to
\iy$. Then $p_{n_m}\to p$ strongly as $m\to \iy$, since $f_0$ is a
homeomorphism between $\mH$ and $\mH_0$. Assume $q=p$. Then $f$ is
a local homeomorphism and $f(q_{n_m})=f(p_{n_m}), q_{n_m}\neq
p_{n_m}$ in a small neighborhood of $q$. We have a contradiction.
Thus $q\neq p$ and $S$ is closed. Condition $f^{-1}(0)=0$ yields
$S\neq \mH$, then $S=\es$. \BBox

Recall that the zeros of $\vp_n(\l)$ are real, simple and strictly
interlace those of $\vp_{n+1}(\l)$. Moreover, the zeros of
$\vt_n(\l)$ are real, simple and strictly interlace those of
$\vp_{n}(\l)$. We recall the well known identities
\[
\lb{2} \vt_n(\l,p)=-{a_0\/a_1..a_{n-1}}\l^{n-2}+O(\l^{n-3}),\ \ \
\ \vp_n(\l,p)={\l^{n-1}\/a_1..a_{n-1}}+O(\l^{n-2}),
\]
\[
\lb{3} \D(\l,p)=\l^{N}-{\l^{N-2}\/2}H-O(\l^{N-3})..,\ \ \
H=b^2+2a^2.
\]
Let  ${\pa}=\pa_p$. We need the simple results.

 \no  \begin{lemma}  \lb{L2.1}
The functions $\m_n,\l_n, \x_n\ev |h_n|^2, n\in \N_ {N-1}$ are real
analytic on $\mH^2$ and the following identities are fulfilled:
\[
\lb{n} d_p\m_n=-{\pa \vt_{N+1}(\m_n(p),p)\/\vt_{N+1}'(\m_n,p)}, \ \ \
\
\]
\[
\lb{L}d_p \l_n=-{\pa\D'(\l_n(p),p)\/\D''(\l_n,p)},
\]
\[
\lb{5} d_p\x_n= (-1)^{s_n}{\pa \D(\l_n(p),p) \/(d\cosh
\sqrt{\x_n}/d\x_n)},\ \ \ \ s_n=N-n.
\]
\end{lemma}

\no {\it Proof.} The function $\D'(\l,p)$ is a polynomial of
degree $N-1$ in $\l$, whose coefficients are polynomials in the
components of $p$. Therefore, its roots $\l_n, n\in \N_ {N-1}$ are
continuous functions of $p$. Moreover, these roots are simple,
then they are real analytic on $\mH^2$. To calculate the gradient,
we observe that $\D'(\l_n(p), p)=0.$ Hence
$$
0=d_p\{\D'(\l_n(p),p)\}=\D''(\l_n(p),p)d_p\l_n+ \pa\D'( \l_n,p),
$$
which implies \er{L}. The proof for $\m_n$ is similar. The
functions $\m_n$  and $\D$ are real analytic, then $\x_n$ is so.
The differentiation of  $(-1)^n\D(\l_n(p),p)= \cosh
\sqrt{\x_n(p)}$ yields
$$
(-1)^n(d\cosh \sqrt{\x_n}/d\x_n)d_p\x_n=
\pa\D(\l_n(p),p)+\D'(\l_n(p),p)d_p\l_n(p),
$$
and the identity $\D'(\l_n(p),p)=0$ implies \er{5}. \BBox

We need the result concerning the mapping $h_{n}(\cdot )$.

\no  \begin{lemma}  \lb{L23} Each function $h_{n}(\cdot ),
n\in \N_ {N-1}$, is real analytic on $\mH^2$ and the following
identities hold
\[
\lb{6} d_p h_{1n}={\vt_{N}'(\m_n(p),p)d\m_n+\pa \vt_{N}(\m_n(p),p)
\/\vt_{N}(\m_n(p),p)},
\]
\[
\lb{7} (-1)^{s_n} (\sinh h_{1n})d_ph_{1n}= \pa \D(\m_n(p),p)+
\D'(\n_n(p),p)d_p\m_n.
\]
Moreover, there exists a real analytic and positive function
$\b_n$ on $\mH^2$ such that
\[
\lb{8}
 h_{2n}(p)=\b_n(p)(\l_n(p)-\m_n(p)), \ \ \ all \ \ \ \
 p\in \mH^2.
\]
\end{lemma}

\no  {\it Proof.} By Lemma 2.1, each $h_{1n}$ is real analytic.
Using Lemma 2.1 and differentiating $\log
(-1)^{s_n}\vt_N(\m_n(p),p)$, we obtain \er{6}. The differentiation
of $2\cosh h_{1n}=(-1)^{s_n}\D(\m_n(p),p)$  gives \er{7}.
 Introduce  the function
$f_n(p)=f(\x_n(p),\x_{1n}(p)), p\in \mH^2,$ where
$$
f(x,y)=2\rt({1\/2}+{(x+y)\/4!}+{(x^{2}+xy+y^{2})\/6!}+\dots
+{x^n-y^n\/(2n)!(x-y)}+...\rt),\ \ \ \ \ \x_{1n}=h_{1n}^{2},
\x_{n}=h_{n}^{2}.
$$
 $f$ is an entire function of two parameters $x,y\in \C$ and $f(x,y)>0$
if $x\ge 0, y\ge 0.$ Then $f_n(\cdot)$ is a real analytic and
positive  on $\mH^2.$  Let $\m_n=\m_n(p), \l_n=\l_n(p)$. Fixing $p\in \mH^2,$ we apply Taylor's
formula for $\D(\l,p)$ with remainder in integral form, at
$\l=\l_n:$
$$
0\le (-1)^{s_n}(\D(\l_n,p)-\D(\m_n,p))=\t_n^{2}g_n(p)/2,\ \ \
\t_n=\m_n-\l_n,
$$
$$
g_n(p)\ev (-1)^{s_n+1}\lt(\D''(\l_n,p)+\t_n(p)\int
_0^1(1-t)^{2}\D'''(\l_n+ t\t_n(p),p)dt\rt).
$$
Using the properties of $\D, \m_n, \l_n $ we get that the function
$g_n$ is real analytic and positive on $\mH^2$. Introduce the real
analytic and positive function $y_n=g_n(p)/f_n(p)$ on $\mH^2$.
Then we can define a real analytic and positive function $\b
_n=\sqrt{y_n}$ on $\mH^2$.  Identities \er{a7},\er{i} yield
$$
\t_n^2g_n/2=(-1)^{n}(\D(\l_n,\cdot)-\D(\m_n,\cdot))= \cosh \sqrt{\x_n} -\cosh
\sqrt{\x_{1n}} =(\x_n-\x_{1n})f_n/2.
$$
 Therefore, we obtain $\x_n-\x_{1n}=\t_n^{2}y_n,$ which implies \er{8}, indeed:
$$
h_{2n}(p)=|\x_n(p)-\x_{1n}(p)|^{1/2}{\rm sign}(\l_n(p)-\m_n(p))= (\l
_n(p)-\m_n(p))\b_n(p), \qqq p\in \mH^2,
$$
therefore, the function $h_{2n}$ is real analytic on $\mH^2$.
 $\BBox$

 We prove the main theorem.

\noindent  {\bf Proof of Theorem 1.1.}
 We check all conditions of Theorem A for $h:\mH^2\to \R^{2N-2}$.

We check i). By Lemma \ref{L23}, the function $h(\cdot )$ is real
analytic on $\mH^2$. We prove  by
contradiction that the operator $d_ph$ is invertible. 
Let  a vector $g\in \mH^2,
g\neq 0$ be a solution of the equation
\[
\lb{eq} (d_ph)g=0,\qq  {\rm or} \qqq   \{ <d_ph_n,g>=0,
\qq
n\in \N_ {N-1}\},
\]
for some fixed $p\in\mH^2$,
 where $<p,\wt p>=\sum_1^{N}(x_n\wt
x_n+b_n\wt b_n)$ is the inner product in $\R^{2N}$. We introduce
the polynomial $f(\l)\ev<(\pa \D)(\l,p),g>,  \l\in \C$,
 of degree $N-2$ with respect to $\l$ (see \er{3}).
 The function $\x_n=h_{1n}^2+h_{2n}^2$ is
analytic  and \er{eq} implies $<d_p\x_n,g>=0.$  Then \er{5} yields
$$
 f(\l_n)=(-1)^{s_n}{d\cosh \sqrt{\x_n}\/d\x_n} <d_p\x_n,g>=0,\ \ \ \ \
 {\rm all} \ \ \ \ n\in \N_ {N-1},
$$
which gives $f\ev 0$. For fixed $p\in \mH^2$ we have 3 cases:

\no 1) Let $h_{2n}=0.$ The differentiation of \er{8} yields
\[ \lb{9}
d_vh_{2n}=\b_n(v)(d_v\l_n(v)-d_v\m_n(v)), \ \ \ {\rm if}\ \ \
h_{2n}(v)=0 \ \ \ {\rm for\  some} \ \ v\in \mH^2.
\]
Then using \er{L},\er{eq} and $f=0$, we obtain $<d_p\l_n,g>=0$ and therefore $<d_p\m_n,g>=0$.

\no 2) Let $h_{1n}\neq 0, h_{2n}\neq 0,$. Then \er{1d}, \er{a7}
yield $\l_n\neq \m_n$. Thus identity \er{7} and $f=0$  imply
$$
0=(-1)^n \sinh h_{1n}<d_qh_{1n},g>=\D'(\m_n(p),p)<d_p\m_n,g>
$$
and then we have  $<d_p\m_n,g>=0,$ since $f\ev 0$ and
$\D'(\m_n)\neq 0.$

\no 3) Let $h_{1n}=0\neq h_{2n}$. Using \er{7} we have
$\pa \D(\m_n(p),p)=-\D'(\m_n(p),p)d_p\m_n$. Identity \er{i} gives
$\vt_N(\m_n,p)=\vp_{N+1}(\m_n,p)=(-1)^{s_n}$, then $\l_n\neq \m_n$
and $\D'(\m_n(p),p)\neq 0$. Thus due to $f=0$ we have $<d_p\m_n,
g>=0$.

Assume that the
vectors $\{d_qh_{1n}, d_p\m_{n}\}_1^{N-1}$ form a basis of
$\mH^2$.
 Then $g=0$ and the operator $d_ph$ is invertible.

Using  standard arguments (see [PT]), we will show that $\{d_p\m_n, d_ph_{1n}\}_1^{N-1}$ is a basis for
$\mH^2$. Recall the result of van
Moerbeke (77p.,[vM]): for each $p\in \mH^2$ the following
identities hold: $\{\m_n,h_{1m}\}_1=\d_{n,m},
\{h_{1n},h_{1m}\}_1=0, \{\m_n,\m_{m}\}_1=0$, where
$\{F,G\}_1=<Jd_pF,d_pG>$ is the Poisson bracket between two
functions $F,G$, for some matrix $J^*=-J$. This gives that
$\{d_p\m_n, d_ph_{1n}\}_1^{N-1}$ is a basis for $\mH^2$. Indeed,
assume that they are linearly dependent, i.e.,
$0=\sum_1^{N-1}(\a_nd_p\m_n+\b_nd_ph_{1n})$ for some $(\a_n,\b_n)_1^{N-1}\in \R^{2N-2}\sm \{0\}$. If $\a_k\neq 0 $, then using
the result of van Moerbeke, we obtain
$0=\{\sum_1^N(\a_nd\m_n+\b_ndh_{1n}),dh_{1k}\}_1=\a_k$, which yields
contradiction. The proof for other cases is similar.

Condition ii) is simple, since the dimensions of our spaces are
finite.

In order to show Condition iii) we prove estimates  \er{e}. Recall
the estimate from [KoK]
\[
\lb{10} 1\le{c\/2}\le e^{h_+}\le 2c,\ \ \ c={\l_N^--\l_0^+\/2},\ \
c_0={\l_N^-+\l_0^+\/2},\ \ h_+=\max_n{|h_n|}.
\]
Define the vector $\wt\l=(\l_0^+,\l_1^-,\l_1^+,..,\l_N^-)\in
\R^{2N}$. Using $\D(\l,p)^2-4=\prod_{1,\pm}^N
 (\l-\l_n^\pm)$ and \er{3} we have
$$
 \D(\l,p)^2-4=\l^{2N}-H\l^{2N-2}+O(\l^{N-3}),\ \
H=b^2+2a^2=-\sum_{1\le n<m}^N\l_n^\pm\l_m^\pm,\ \
\sum_1^N\l_n^\pm=0.
$$
Note that $\l_0^+<0$ and $\l_N^->0$  since $\sum_1^N\l_n^\pm=0$.
The identity $0=(\sum_1^N\l_n^\pm)^2=\|\wt \l\|^2-2H$ gives
$2H=\|\wt \l\|^2$. Thus the identity
${\l_0^+}^2+{\l_N^-}^2=2(c^2+c_0^2)$ and the estimate $|c_0|<c$
yield
\[
\lb{12} 2c^2<\|\wt \l\|^2=2H<8Nc^2.
\]
Then \er{10} implies \er{e}. Let $x_n^+=\max \{0,x_n\},
x^+=\{x_n^+\}_1^N$ and $x=x^++x^-$. Let $\|x\|_1=\sum |x_n|$ and
using the identity $\|x^+\|_1=\|x^-\|_1$ we get
$e^{2\|x\|_1/N}<a^2$. The last estimate and \er{e} implies
$b^2+2e^{2\|x\|/N}\le 32Ne^{2h_+}$, which yields the estimate
$\|p\|$ in terms of $\|h\|$.

 If $h=0$, then \er{10} gives $c=2$. Recall that $c=2$ iff $p=0$ (see
[KoK]).

 Therefore, all conditions of Theorem
A are fulfilled and $h$ is a real analytic isomorphism between
$\mH^2$ and $\R^{2N-2}$. \BBox

\noindent {\bf Acknowledgments.}\ Evgeny Korotyaev was partly supported by DFG project BR691/23-1.

\noindent {\bf References}
\small

\no  [BGGK]  B\"attig, D.; Grebert, B.; Guillot, J.; Kappeler,
T. Fibration of the phase space of the periodic Toda
   lattice. J. Math. Pures Appl. (9) 72 (1993), no. 6, 553--565.

\no [GT] Garnett J.; Trubowitz E. Gaps and bands of one dimensional
 periodic Schr\"odinger operators. Comment. Math. Helv. 59(1984), 258-312.

\no [KK] Kargaev P.; Korotyaev E. Inverse Problem for the Hill
Operator, the Direct Approach.  Invent. Math., 129(1997), no. 3,
567-593.

\no [K] Korotyaev, E. The inverse problem for the Hill operator.
I. Int. Math. Res. Not. 3(1997), 113--125.

\no [K1] Korotyaev, E. Characterization of the spectrum of Schr\"odinger operators with periodic distributions. Int. Math. Res. Not. 2003, no. 37, 2019--2031.

\no [K2] Korotyaev, E. Gap-Length Mapping for Periodic Jacobi Matrices,
 Russ. J. Math. Phys. 13(2006), no. 1, 64-69.

\no [KKu] 
Korotyaev, E.; Kutsenko, A. Inverse problem for the discrete 1D Schr\"odinger operator with small periodic potentials. Comm. Math. Phys. 261 (2006), no. 3, 673--692.

\no [KoK]  Korotyaev, E.; Krasovsky, I. Spectral estimates for
periodic Jacobi matrices. Comm. Math. Phys. 234 (2003), no. 3,
517--532.

\no [vM] van Moerbeke, P. The spectrum of Jacobi matrices.
Invent. Math. 37 (1976), no. 1, 45--81

\no [MO]  Marchenko V.; Ostrovski I. A characterization of the
spectrum of the Hill operator. Math. USSR Sb. 26(1975), 493-554.

\no [Pe] Perkolab, L. An inverse problem for a periodic Jacobi
matrix. (Russian) Teor. Funktsii Funktsional. Anal. i
   Prilozhen. 42(1984), 107-121.

\no [PT] P\"oschel, J.; Trubowitz, E. Inverse spectral theory.
Pure and Applied Mathematics, 130. Academic Press, Inc., Boston, MA, 1987.

\no [Te] Teschl, G. Jacobi operators and completely integrable
nonlinear lattices. Mathematical Surveys
   and Monographs, 72. American Mathematical Society,
   Providence, RI, 2000.

\end{document}